\documentclass{INTERSPEECH2023}

\usepackage{multirow, tabularx, makecell}
\usepackage[dvipsnames]{xcolor}

\interspeechcameraready

\title{One-Step Knowledge Distillation and Fine-Tuning in Using Large Pre-Trained Self-Supervised Learning Models for Speaker Verification}
\name{Jungwoo Heo$^*$\thanks{$^*$Equal contribution}, Chan-yeong Lim$^*$, Ju-ho Kim, Hyun-seo Shin, and Ha-Jin Yu\sthanks{$^\dag$Corresponding author.}}
\address{School of Computer Science, University of Seoul}
\email{jungwoo4021@gmail.com, cksdud585@naver.com, wngh1187@naver.com, gustjtls123@naver.com, hjyu@uos.ac.kr}

\begin{document}

\maketitle
 
\begin{abstract}
The application of speech self-supervised learning (SSL) models has achieved remarkable performance in speaker verification (SV). 
However, there is a computational cost hurdle in employing them, which makes development and deployment difficult. 
Several studies have simply compressed SSL models through knowledge distillation (KD) without considering the target task. 
Consequently, these methods could not extract SV-tailored features. 
This paper suggests One-Step Knowledge Distillation and Fine-Tuning (OS-KDFT), which incorporates KD and fine-tuning (FT). 
We optimize a student model for SV during KD training to avert the distillation of inappropriate information for the SV. 
OS-KDFT could downsize Wav2Vec 2.0 based ECAPA-TDNN size by approximately 76.2\%, and reduce the SSL model's inference time by 79\% while presenting an EER of 0.98\%. 
The proposed OS-KDFT is validated across VoxCeleb1 and VoxCeleb2 datasets and W2V2 and HuBERT SSL models. 
Experiments are available on our GitHub \footnote{\url{https://github.com/jungwoo4021/OS-KDFT}}. 
\end{abstract}
\noindent\textbf{Index Terms}: Speaker verification, Self-supervised learning model, Knowledge-distillation

\section{Introduction}
Speaker verification (SV) verifies whether the input utterance is vocalized by the target speaker. 
Most SV studies have employed hand-crafted acoustic features such as spectrogram, Mel spectrogram, and the Mel-Frequency Cepstral Coefficient (MFCC) as inputs \cite{liu2022learnable, shim2022graph}. 
Recently, in speech signal processing fields, there has been an increasing interest in speech self-supervised learning (SSL) models such as Wav2Vec 2.0 (W2V2) \cite{baevski2020wav2vec}, HuBERT \cite{hsu2021hubert}, and WavLM \cite{chen2022wavlm} because they have the potential to extract more affluent representation than the hand-craft method \cite{hussain2022multi, song2023exploring}. 
Following this trend, a recent SV work achieved remarkable performance using speech SSL models \cite{chen2022large}. 

Despite speech SSL models' impressive performance, there is a computational cost hurdle in employing them. 
The largest version of W2V2 and HuBERT contains approximately 317M and 964M parameters, respectively. 
Because of their large size, the development and deployment of models present a significant challenge. 
Consequently, research communities have focused on building lightweight SSL models, and some have attempted to train small-size models \cite{chi2021audio}. 
However, due to their limited capacity, training a small model with a significant amount of data is difficult \cite{sun2017revisiting, huang2019gpipe}. 
As an alternative approach, several studies have explored the possibility of knowledge distillation (KD), which is a well-known model compression strategy that transfers knowledge from a large model to a smaller one \cite{hinton2015distilling}. 
Sanh \textit{et al.} \cite{sanh2019distilbert} and Jiao \textit{et al.} \cite{DBLP:conf/emnlp/JiaoYSJCL0L20} have devised DistilBERT and Tinybert that distills the BERT and demonstrates the potential of KD in the natural language processing (NLP). 
Following these approaches, in the acoustic signal processing field, Lee \textit{et al.} \cite{DBLP:conf/interspeech/LeeJGJK22} and Chang \textit{et al.} \cite{chang2022distilhubert} designed FitHuBERT and DistilHuBERT to deliver outstanding performance on the speech processing universal performance benchmark \cite{yang2021superb}. 
Moreover, Peng \textit{et al.} successfully downsized W2V2 framework through KD \cite{peng2021shrinking}. 
Therefore, studies using KD have become mainstream in current research.

\begin{figure}[t]
\begin{center}
    \centering
    \includegraphics[width=0.85\linewidth]{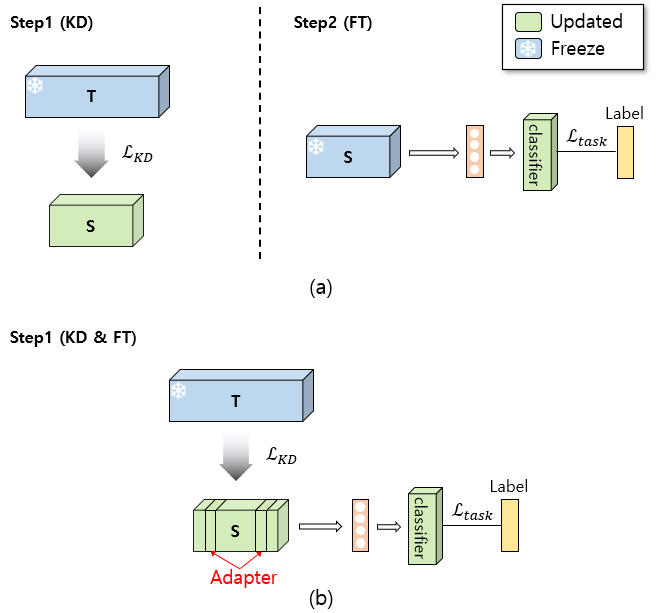}
    \vspace{-0.2cm}
    \caption{
    Comparison of the previous SSL model KD and proposed OS-KDFT. 
    (a) describes the training steps of the previous method that trains KD and FT independently. 
    (b) illustrates the process of OS-KDFT, which perform KD and FT simultaneously. 
    Model `\textbf{T}' and `\textbf{S}' means 'teacher network' and 'student network'. 
    }
\label{figure:overall}
\end{center}
\vspace{-3em}
\end{figure}

Nevertheless, simply compressing SSL models through KD has fundamental limitations. 
Large SSL models are often optimized for the target task to extract task-tailored features, which are better than the original features \cite{hussain2022multi, song2023exploring, chen2022large}. 
As illustrated in Figure \ref{figure:overall} (a), previous SSL compressing studies first constructed a lightweight SSL model through KD and then utilized it for the target task with a fixed state. 
Because they cannot consider the target task, they could not extract the task-customized features. 
Fine-tuning (FT) of the distilled SSL models can mitigate this concern, it demands additional training. 
Furthermore, determining the appropriate transition point from KD to FT is difficult because knowing the optimal quantity of KD for FT is complicated. 
Thus, iterative empirical experiments may be required to decide the best transition point. 

In this paper, we aim to compress the SSL model for SV tasks efficiently to facilitate development and deployment. 
We proposed a novel training strategy One-Step Knowledge Distillation and Fine-Tuning (OS-KDFT) that jointly trains KD and FT. 
We believed that by performing KD and FT concurrently, teacher networks could effectively transfer crucial information for SV. 
Therefore, in the proposed OS-KDFT, we incorporate KD and FT training processes, as depicted in Figure \ref{figure:overall} (b). 
Through this, the proposed method can perform KD while considering the target task to avert the distillation of inefficient information for the target task. 
Moreover, this method avoids the picking transition point of KD and FT learning. 
OS-KDFT is explained in detail in Section 3. 

Through this paper, we make the following contributions. 
\begin{itemize}\setlength{\leftskip}{15pt}
    \item To the best of our knowledge, OS-KDFT is the first approach to compress the speech SSL model while concurrently fine-tuning. 
    Previous studies have only concentrated on condensing the SSL through KD, but we also considers FT to extract task-tailored features. 
    \item The proposed OS-KDFT has effectively reduced W2V2 + ECAPA-TDNN size by 76\%, and the SSL model's inference time by 79\% while showing an EER of 0.98\%. 
    \item The proposed OS-KDFT is validated across VoxCeleb1 and VoxCeleb2 datasets and W2V2 and HuBERT speech SSL models. 
\end{itemize}

\section{Related work}
Speech SSL models have demonstrated satisfactory performance in many acoustic signal processing research. 
Nevertheless, due to their high computational costs, studies on model compression have drawn attention \cite{sanh2019distilbert, DBLP:conf/emnlp/JiaoYSJCL0L20}. 
This section introduces \textit{i)} previous efforts, \textit{ii)} why we studied KD, and \textit{iii)} the research in the NLP that inspired our proposed method. 

\subsection{SSL model compression}
Model compression methods include quantization, pruning, and knowledge distillation. 
To reduce the size of each parameter, quantization represents weights in lower bitwidth representations. 
Wang \textit{et al.} \cite{wang2022deep} suggested the potential of quantization by successfully reducing the W2V2 framework. 
Nevertheless, quantization has a limitation in that it cannot reduce the number of parameters. 
Pruning removes weights or channels that have minimal impact on the output \cite{NIPS2015_ae0eb3ee}. 
Lai \textit{et al.} \cite{lai2021parp} showed the superiority of this technique by applying the pruning method to the speech SSL model. 
However, pruning presents difficulties in establishing appropriate criteria for selecting the parameters to be pruned. 
Knowledge distillation refines knowledge from a large model to a small model. 
KD could reduce the number of parameters and avoid the effort of finding parameters to remove. 
Furthermore, this method demonstrated its effectiveness in various lightweight speech SSL frameworks, such as FitHuBERT \cite{DBLP:conf/interspeech/LeeJGJK22}, Distilhubert \cite{chang2022distilhubert}, and LightHuBERT \cite{DBLP:conf/interspeech/WangBAZXWZKL22}. 
Therefore, in following this trend, we utilized KD as a technique for compressing the speech SSL model. 
 
\subsection{Distribution mismatch between teacher and student}
Researchers in the NLP have argued that the ideal distribution for students might be different from the teacher's output \cite{jin2019knowledge, park2021learning}. 
Their studies determined that distribution disparity can occur despite teachers and students conducting the identical task. 
In our study, the student model learns SV that the teacher model has never trained. 
Therefore, the distributional gap between the ideal student and teacher might be more significant than the NLP research. 
From this perspective, we modified the architecture of the student network to bridge distribution mismatch as in Section 3.

\begin{figure}[t]
\begin{center}
    \centering
    \includegraphics[width=0.93\linewidth]{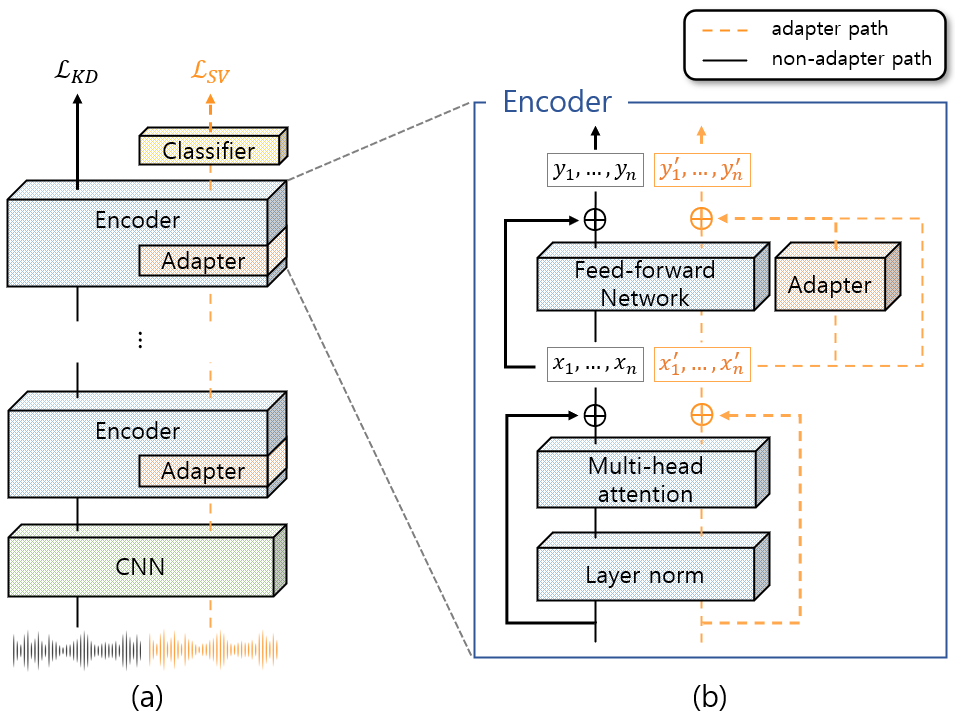}
    \vspace{-0.2cm}
    \caption{
    The architecture of the student network (a) and encoder module (b). 
    There are two routes, a non-adapter path and an adapter path depending on whether they go through adapters or not. 
    The features passing through the non-adapter path are used for KD training. 
    On the other hand, the adapter path indicates the path of the mini-batch utilized for performing SV. 
    ($\oplus$ : element-wise summation.) 
    }
\label{figure:encoder_structure}
\end{center}
\vspace{-3em}
\end{figure}

\section{OS-KDFT}
Model compression has become an increasingly important topic in the research community, because it can facilitate the development and deployment of deep neural networks. 
Many SSL model application studies have explored various KD techniques, but they only focused on distilling teachers' knowledge regardless of the target tasks. 
Thus, this study aims to transfer teachers' knowledge to be suitable for SV. 
To this end, we suggest the novel training strategy OS-KDFT, which incorporates teacher knowledge transfer and speaker verification tuning. 
OS-KDFT employs a unique structured student network that has two branches to imitate teachers' output and perform the target task. 
This section explains the overall architecture of the student, the weight initialization, and the learning rate setting, modified from the original. 

\subsection{Architecture}
We distilled the W2V2 and HuBERT frameworks, consisting of convolutional neural network (CNN) and transformer encoders. 
We constructed a student network with reduced number of parameters by limiting the number of transformer encoder layers. 
Figure \ref{figure:encoder_structure} (a) describes the architecture of the student network, in which the student network has two routes: the non-adapter path for KD and the adapter path for SV. 
Both paths share CNN and encoder weights, but only the adapter path utilizes independent parameters via adapters. 
We divided the student's branches to mitigate negative transfer, which is performance degradation due to conflicts between tasks in a multi-task setting \cite{liu2019loss}. 
Because we jointly optimized the student network with two different tasks (KD and FT), the student was exposed to negative transfer. 
One of the well-known solutions for mitigating the negative transfer is using independent parameters for each task \cite{meng2021multi, sun2020adashare}. 
Thus, we divided the student's branches and inserted additional parameters (adapters, inspired by LORA \cite{DBLP:conf/iclr/HuSWALWWC22}) into the one performing SV. 

The detailed process of the encoder block is depicted in Figure \ref{figure:encoder_structure} (b) and Equations (1)–(3). 
In the non-adapter path, the \textit{Layer norm} and \textit{Multi-head attention} layers computes hidden features $X=\{x_1, x_2, ..., x_n\}$. 
After that, $x_i$ is converted to $f(x_i)$ through the feed-forward layer (Equation 1). 
The output $Y=\{y_1, y_2, ..., y_n\}$ is the element-wise summation of $x_i$ and $f(x_i)$, as in Equation (2). 
In the adapter path, the hidden features $X'=\{x'_1, x'_2, ..., x'_n\}$ is calculated via the \textit{Layer norm} and \textit{Multi-head attention} layers, and $x'_i$ is transformed to $f(x'_i)$ with the identical mechanism in the non-adapter path. 
The $x'_i$ is also fed to the adapter, which consists of downsampling ($W_{down} \in \mathbb{R}^{1024 \times 64}$), a ReLU activation function, and upsampling ($W_{up} \in \mathbb{R}^{64 \times 1024}$). 
As explained by Equation (3), the out features $y'_i$ are the element-wise summation of $x'_i$, $f(x'_i)$, and the output of the adapter. 
\begin{gather}
    f(x_i) = Feed \thinspace Forward(x_i) \\
    y_i =  x_i + f(x_i) \\
    y'_i = x'_i + f(x'_i) + ReLU(x'_iW_{down})W_{up} 
\end{gather}

\subsection{Weight initialization \& learning rate}
We randomly initialize the parameters of classifiers and adapters because the original SSL model (teacher) does not contain both modules. 
On the other hand, CNN and encoder weights are generated utilizing teacher weights. 
When initializing the student encoders, we use the weights from the teacher encoders in order of closeness to the CNN. 
This strategy is based on Chen \textit{et al.}'s findings that an encoder closer to a CNN can extract affluent features for SV \cite{chen2022large}. 

Chen \textit{et al.} froze the W2V2 during the first 10 epochs to alleviate the disparity of learning quantity between W2V2 and ECAPA-TDNN. 
By imitating their strategy, we adjusted different learning rates on the CNN, encoders, adapters, and classifiers. 
Equations (4)–(6) describe the learning rate of each module in epoch $\tau$. 
The learning rate of the randomly initialized classifier ($lr_{c}^{\tau}$) was reduced following a cosine annealing rule (Equation 4). 
On the other hand, the learning rate of the pre-trained CNN and encoders ($lr_{s}^{\tau}$) gradually increased during the initial 10 epochs and then decreased (Equation 5). 
In the adapters, the learning rate ($lr_{a}^{\tau}$) decreased from the maximum to the minimum since it was also randomly initialized. 
But the value was adjusted by multiplying by $\theta$ (Equation 6). 
We set $\beta$ to 0.93 and $\theta$ to 10 because they delivered the best results in our experiments. 
\begin{gather}
    lr_{c}^{\tau} = \eta_{min} + \frac{1}{2}(\eta_{max}-\eta_{min})(1+cos(\frac{\tau}{\tau_{tot}}\pi)) \\
    lr_{s}^{\tau} =
    \begin{cases}
    lr_{c}^{\tau} \times \frac{\tau}{10} ,& \tau \leq 10 \\
    lr_{s}^{\tau - 1} \times \beta ,& \tau > 10
    \end{cases} \\
    lr_{a}^{\tau} = lr_{c}^{\tau} \times \theta
\end{gather}

\section{Experiment setting}

\subsection{Dataset}
We used VoxCeleb1 \cite{nagrani2017voxceleb} and VoxCeleb2 \cite{chung2018voxceleb2} datasets to evaluate our proposed method. 
The VoxCeleb1 training set is comprised of 148,642 utterances from 1,211 speakers, and the test set consists of 4,874 utterances from 40 speakers. 
The VoxCeleb2 training set corresponds to 1,092,009 samples that were collected from 5,994 speakers. 
We only used the VoxCeleb2 training set without a test partition. 
For the data augmentation (DA), we employed MUSAN \cite{snyder2015musan} and RIR reverberation \cite{ko2017study} datasets. 
We evaluated the models using all three official VoxCeleb1 trials: Vox1-O, Vox1-Extend, Vox1-Hard. 
The primary metric is the equal error rate (EER) based on cosine similarity.

\begin{table}[t]
\caption{
Comparison of the equal error rate (\%) and SSL model's inference time between W2V2 + ECAPA-TDNN and compressed models, generated through KDFT and proposed OS-KDFT, on the VoxCeleb2 dataset (*: our implementation). 
The inference time was measured on an NVIDIA RTX A5000 GPU with a batch size of one. 
We repeated inference time measurement 100 times and recorded the average value. 
}
\centering
\resizebox{\linewidth}{!}{%
\label{table:main}
\begin{tabular}{lccc}
\Xhline{2\arrayrulewidth}
\textbf{Experiment} & \textbf{Size} & \textbf{Inf. time} & \textbf{EER (\%)} \\
\hline
\#1 W2V2 + ECAPA-TDNN (small) \cite{chen2022large} & \multirow{2}{*}{321.4M} & N/A & 0.73 \\
\#2 W2V2 + ECAPA-TDNN (small)* & & 48.8ms & 0.82 \\
\hline
\#3 KDFT & 76.1M & 10.5ms & 1.26 \\
\#4 OS-KDFT & 76.6M & 10.5ms & 0.98 \\
\Xhline{2\arrayrulewidth}
\end{tabular}}
\vspace{-1em}
\end{table}

\subsection{Baseline}
Based on Chen \textit{et al.}, we defined a framework combining the speech SSL model and ECAPA-TDNN as the baseline \cite{chen2022large}. 
We implemented the baseline using the HuggingFace \cite{wolf-etal-2020-transformers} transformers library and exploited pre-trained W2V2 and the HuBERT version of XLSR \footnote{facebook/wav2vec2-large-xlsr-53} and large\footnote{facebook/hubert-large-ll60k}. 

\subsection{Experiment details}
We constructed a mini-batch using 128 samples, and each sample was randomly cropped into a 2-second length. 
Then, we employed the Adam optimizer without weight decay, utilized the mean-squared error (MSE) for KD learning, and multiplied it by 100 to adjust the ratio between losses. 
As the speaker classification loss function, we used the AAM-softmax \cite{deng2019arcface} criterion with a margin of 0.15 and a scale of 20. 
We applied SpecAugment on the output of SSL model in our experiment with data augmentation. 
In the evaluation, the entire utterance and its five segments of 3 seconds were utilized. 
Further details can be found on our GitHub.

\section{Results}
\begin{figure}[t]
\begin{center}
    \centering
    \includegraphics[width=0.9\linewidth]{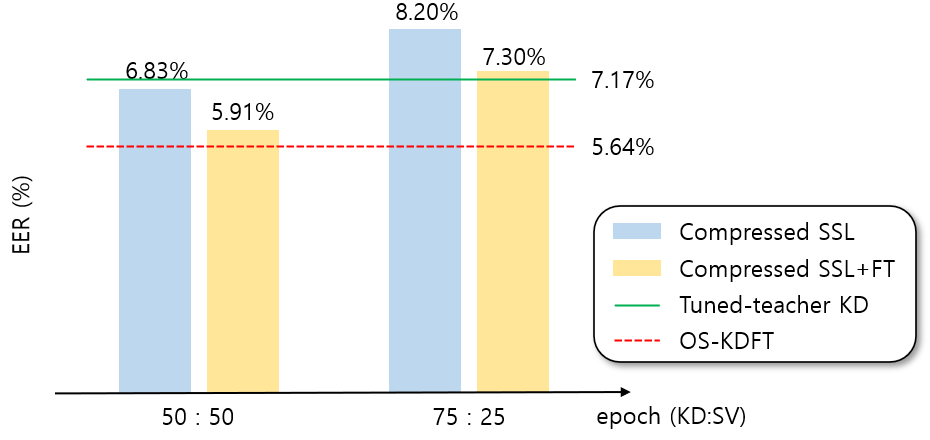}
    \vspace{-0.2cm}
    \caption{Comparison of SSL model KD, tuned-teacher SSL KD, and OS-KDFT in VoxCeleb1. 
    To exclude variables, we did not apply DA. 
    The \textcolor{cyan}{blue (left)} and \textcolor{yellow}{yellow (right)} bars display the EER of the compressed SSL model for frozen and FT versions, respectively. 
    The \textcolor{green}{green solid line} and \textcolor{red}{red dotted line} results from tuned-teacher KD and OS-KDFT. 
    The x-axis represents the ratio of epochs for KD and SV.}
\label{figure:twostep}
\end{center}
\vspace{-3em}
\end{figure}

\textbf{Comparison with the baseline.} 
Table \ref{table:main} compares conventional frameworks and the proposed OS-KDFT. 
Experiment \#1 is the baseline framework present in \cite{chen2022large}, and Experiment \#2 is our implementation. 
Each experiment demonstrates an EER of 0.73\% and 0.82\% without score calibration. 
Experiments \#3 and \#4 are the results of compressing the framework of Experiment \#2 through knowledge distillation and fine-tuning (KDFT) and OS-KDFT, respectively. 
KDFT is a training strategy that incorporates KD and SV learning without any modification, and OS-KDFT is our proposed method that developed KDFT using adapters. 
KDFT and OS-KDFT significantly compress the baseline by approximately 76\% and reduce the SSL model's inference time by 79\%. 
KDFT achieves an EER of 1.26\%, a seriously degraded performance from 0.82\% (\#3). 
However, OS-KDFT successfully carries out KD and FT and delivers an EER of 0.98\% (\#4). 
Through these results, we confirm that the proposed OS-KDFT has the potential to distill the SSL model suitable for SV. 
\\\\
\textbf{Comparison with conventional method.}
To investigate further, we compared OS-KDFT with other training strategies in the VoxCeleb1 dataset; Figure \ref{figure:twostep} illustrates these results. 
In these experiments, we did not apply DA to exclude variables. 
The blue (left) bars in Figure \ref{figure:twostep} show the results of compressing the SSL model via KD and its use for SV. 
This method achieved EER of 6.83\% and 8.20\% when the epoch ratio of KD and FT was at 50:50 and 75:25, respectively. 
The yellow (right) bars depict the result of further tuning for SV: this decreased the EER to 5.91\% and 7.30\%, respectively for each epoch ratios. 
These results confirm that simply compressing the model with KD does not generate optimal students for the target task. 
In addition, performance deviations may occur depending on the proportion of learning KD and SV. 
The green solid line represents the EER of the student that acquired knowledge from the SV-tuned teacher. 
Since the teacher can identify speakers, we could utilize Kullback-Leibler divergence loss for student training in this experiment. 
As a result, we trained the student to predict teachers' softmax output distribution, resulting in the EER of 7.17\%. 
The red dotted line represents the OS-KDFT results, and it delivers the lowest EER of 5.91\%. 
These results show that a KD and FT joint training strategy has potential compared to conventional compression methods. 
\\\\
\textbf{Ablation study.} 
Table \ref{table:abal1} displays the performance variation with an application of each strategy of OS-KDFT on the VoxCeleb1 and VoxCeleb2 datasets. 
Compressing the baseline through KDFT in VoxCeleb1 dataset yields an EER of 8.17\% (\#5). 
In Experiment \#6, we added only the adapter's parameters to \#5 without separating branches, and it demonstrated a marginally improved EER of 7.94\%. 
When dividing routes (\#7), it achieves the EER of 7.28\%, which is a significantly enhanced performance than \#5. 
Finally, we can reach the best performance of 5.64\% by adjusting the learning rate described in Section 3.2. 
In experiments using VoxCeleb2, we used not only \textit{Original (O)} but \textit{Extend (E)} and \textit{Hard (H)} trials for a more sophisticated evaluation. 
When the baseline was compressed through KDFT, it delivered EERs of 3.35\%, 3.41\%, and 5.97\% in O, E, and H trials, respectively (\#9). 
In Experiment \#10, we simply add the adapter's parameters without branch split, and it degrades the performance to 3.46\%, 3.41\%, and 6.01\%. 
On the other hand, in experiment \#11, we also divide the path and improve EER to 2.74\%, 3.01\%, and 5.44\%. 
The best performances of 2.50\%, 2.56\%, and 5.18\% are delivered when the learning rate is also diversified for each module (\#12). 
Therefore, it is difficult to attribute the effect of OS-KDFT to simply increasing the number of parameters and each strategy of OS-KDFT is necessary.  
\\\\
\textbf{Application to other models.}
To further verify the effectiveness of OS-KDFT, we applied this to another SSL model and classifier. 
Table \ref{table:various_framework} presents the results of these studies. 
In Experiments \#13 and \#14, we changed the classifier from ECAPA-TDNN to a linear layer. 
The model trained identically to \#3 delivered an EER of 8.27\%, while the framework trained with OS-KDFT presented an EER of 5.92\%. 
In Experiments \#15 and \#16, we used HuBERT instead of W2V2. 
Distilling HuBERT through KDFT offered an inferior EER of 7.15\%. 
In contrast, when compressing HuBERT via OS-KDFT, the EER was 5.97\%. 
Through these results, we can confirm that the OS-KDFT method effectively works in other frameworks.

\begin{table}[t]
\caption{
Equal error rate (\%) for different training strategies which are trained by VoxCeleb1(\#5-8) \& VoxCeleb2(\#8-12). 
O, E, and H are official trial lists Vox1-O, Vox1-E, and Vox1-H, respectively. 
AS indicates that adapters are inserted into the student encoder layer, and LR means we set different learning rates as described in Section 3.2. 
}
\centering
\vspace{-0.1cm}
\resizebox{\linewidth}{!}{%
\label{table:abal1}
\begin{tabular}{clcccc}
\Xhline{2\arrayrulewidth}
\multicolumn{5}{l}{\textbf{VoxCeleb1 train (without DA)}} \\
\textbf{\#Exp} & \textbf{Train strategy} & \textbf{Size} & \textbf{EER (O)} & \textbf{EER (E)} & \textbf{EER (H)} \\
\hline 
\#5 & KDFT & 76.1M & 8.17 & N/A & N/A \\
\#6 & KDFT (AS param) & 76.6M & 7.94 & N/A & N/A \\
\#7 & OS-KDFT (AS) & 76.6M & 7.28 & N/A & N/A \\
\#8 & OS-KDFT (AS, LR) & 76.6M & \textbf{5.64} & N/A & N/A \\
\hline \\
\Xhline{2\arrayrulewidth} 
\multicolumn{5}{l}{\textbf{VoxCeleb2 train (without DA)}} \\
\textbf{\#Exp} & \textbf{Train strategy}  & \textbf{Size} & \textbf{EER (O)} & \textbf{EER (E)} & \textbf{EER (H)} \\
\hline 
\#9 & KDFT & 76.1M & 3.35 & 3.41 & 5.97 \\
\#10 & KDFT (AS param) & 76.6M & 3.46 & 3.41 & 6.01 \\
\#11 & OS-KDFT (AS) & 76.6M & 2.74 & 3.01 & 5.44 \\
\#12 & OS-KDFT (AS, LR) & 76.6M & \textbf{2.50} & \textbf{2.56} & \textbf{5.18} \\
\Xhline{2\arrayrulewidth}
\end{tabular}}
\vspace{-0.5em}
\end{table}

\begin{table}[t]
\caption{
    Experimental results with different PLMs and classifier. 
    Experiments were performed on the VoxCeleb1 dataset, and data augmentation was not applied. 
}
\centering
\vspace{-0.1cm}
\resizebox{0.9\linewidth}{!}{%
\label{table:various_framework}
\begin{tabular}{ccccc}
\Xhline{2\arrayrulewidth}
\textbf{\#Exp} & \textbf{Train strategy} & \textbf{SSL model} & \textbf{Classifier} & \textbf{EER (\%)} \\
\hline
\#13 & KDFT & \multirow{2}{*}{W2V2} & \multirow{2}{*}{Linear} & 8.27 \\
\#14 & OS-KDFT & & & \textbf{5.92} \\
\hline
\#15 & KDFT & \multirow{2}{*}{HuBERT} & \multirow{2}{*}{Linear} & 7.15 \\
\#16 & OS-KDFT & & & \textbf{5.97} \\
\Xhline{2\arrayrulewidth}
\end{tabular}}
\vspace{-2em}
\end{table}

\section{Conclusion}
In this paper, we design a One-Step Knowledge Distillation and Fine-Tuning (OS-KDFT) method to condense SSL model for SV. 
OS-KDFT is the first approach to compress the speech SSL model while concurrently fine-tuning and it mitigating negative transfers by utilizing adapters. 
Through OS-KDFT, we can compress the 321.4M model to 76.6M, and reduce the SSL model's inference time by 79\% while presenting an EER of 0.98\%. 
We have verified the effectiveness of the OS-KDFT through comparison with other training strategies and applications on other SSL model. 
Nevertheless, our research has limitations. 
To generalize the effectiveness of OS-KDFT, we should evaluate our proposed method with different KD methods (e.g., transferring teachers' knowledge from hidden features rather than the output). 
Thus, we will incorporate OS-KDFT with different KD methods in future works. 

\section{Acknowledgements}
This work was supported by the National Research Foundation of Korea(NRF) grant funded by the Korea government. (MSIT) (2023R1A2C1005744)

\bibliographystyle{IEEEtran}
\bibliography{refs}

\end{document}